\documentclass[12pt,preprint]{aastex}
\usepackage{emulateapj5}
\usepackage{epsfig}

\newcommand{\etal}{\mbox{et al.}}
\newcommand{\ergcms}{erg cm$^{-2}$ s$^{-1}$}
\newcommand{\ergsec}{erg s$^{-1}$}
\newcommand{\phcms}{photons cm$^{-2}$ s$^{-1}$}

\newcommand{\msun}{$M_{\odot}$}

\newcommand{\forbit}{\mbox{CXOGC~J174532.7$-$290552}}

\newcommand{\chandra}{{\it Chandra}}
\newcommand{\rosat}{{\it ROSAT}}

\newcommand{\asca}{{\it ASCA}}

\newcommand{\sgrastar}{\mbox{Sgr A$^*$}}
\newcommand{\program}[1]{{\tt {#1}}}
\newcommand{\html}[1]{{\tt http://#1}}

\makeatletter
\newenvironment{inlinefigure}{%
\def\@captype{figure}%
\noindent\begin{minipage}{0.999\linewidth}\begin{center}}
{\end{center}\end{minipage}\smallskip}
\makeatother

\shortauthors{Muno \etal}
\shorttitle{Periodic X-ray Sources}

\begin{document}
\title{X-ray Sources with Periodic Variability in a Deep \chandra\ Image of the Galactic Center}
\author{M. P. Muno,\altaffilmark{1} 
F. K. Baganoff,\altaffilmark{1} M. W. Bautz,\altaffilmark{1}
W. N. Brandt,\altaffilmark{2} G. P. Garmire,\altaffilmark{2}
and G. R. Ricker\altaffilmark{1}}

\altaffiltext{1}{Center for Space Research,
Massachusetts Institute of Technology, Cambridge, MA 02139;
muno@space.mit.edu, fkb@space.mit.edu}
\altaffiltext{2}{Department of Astronomy and Astrophysics, 
The Pennsylvania State University, University Park, PA 16802}

\begin{abstract}
We report the discovery of eight X-ray sources with periodic variability 
in 487~ks of observations of the Galactic center with \chandra.
The sources are identified from a sample of 285 objects detected with 
100--4200 net counts. Their
periods range from 300~s to 4.5~h with amplitudes between 40\% and
70\% rms. They have luminosities of $(1 - 5) \times 10^{32}$~\ergsec\ 
(2--8~keV at 8 kpc). The spectra of seven of the eight sources are consistent
with $\Gamma \approx 0$ power laws absorbed by gas and dust with a
column density equal to or higher than that toward the Galactic Center
($6 \times 10^{22}$ cm$^{-2}$). Four of these sources also exhibit emission
lines near 6.7~keV from He-like Fe, with equivalent widths of 600--1000~eV. 
These properties are consistent with both 
magnetically accreting cataclysmic variables and wind-accreting neutron 
stars in high-mass X-ray binaries. The eighth source has an absorbing column 
of $5 \times 10^{21}$ cm$^{-2}$ that places it in the foreground.
Its spectrum is consistent with either a $\Gamma = 1.4$ power law or 
$kT = 25$~keV bremsstrahlung emission. Its period-folded flux profile clearly 
identifies it as an eclipsing polar. We place an approximate upper limit 
of $i^\prime > 23$ magnitude on the optical counterpart to this source using
a 5 min exposure obtained with the MagIC camera on the 
Clay telescope (Magellan II) at Las Campa\~{n}as.
\end{abstract}

\keywords{X-rays: binaries --- stars: rotation --- stars: pulsars --- stars: cataclysmic variables}

\section{Introduction\label{sec:intro}}

In order to understand the star-formation history of our Galaxy, it is
important to know the numbers of stars that have long since ended their 
lives on the main sequence. Although the compact remnants of dead 
stars --- white dwarfs, neutron stars, and black holes --- are generally
difficult to find in isolation, they can be identified relatively easily
if they are accreting significant amounts of matter from a companion star 
in a close binary orbit. As the matter falls in toward the compact object, 
it is heated and shocked, causing it to emit 
$10^{29}$--$10^{39}$~\ergsec
of X-rays (see Warner 1995; White, Nagase, \& Parmar 1995, for reviews).  
When combined with sophisticated models for the evolution of these close 
binaries, the observed numbers of accreting 
X-ray sources can provide estimates of the total numbers of black holes, 
neutron stars, and white dwarfs in the Galaxy 
(Iben, Tutukov, \& Fedorva 1997; Kalogera 1999; Howell, Nelson, \& Rappaport
2001; Podsiadlowski, Rappaport, \& Pfahl 2002)\nocite{itf97,kal99,hnr01,prp02}.

Such a study could be particularly fruitful toward the Galactic center,
where the star-formation history is uncertain \citep{mor93,sm96} 
and the stellar
density is high enough that it is possible to assemble a statistically 
meaningful sample of X-ray sources from a relatively small field.
We have recently carried out a census of X-ray emitting objects brighter than 
$10^{30}$ \ergsec\ in a 17\arcmin\ by 17\arcmin\ field around \sgrastar\
using the {\it Chandra X-ray Observatory} \citep{mun03}. We detected over 2300 
individual point sources in this field, the natures of most of which are
unknown because a wide range of stellar systems emit X-rays at this level.
However, more than half of the 800 sources that were bright enough to 
provide spectral information ($\gtrsim 60$ net counts above 2~keV) had very 
hard spectra, consistent with a 
power law $E^{-\Gamma}$ with photon index $\Gamma < 1$. Such hard spectra 
have only
been observed previously from two classes of source: neutron stars 
accreting from the winds of young, massive companions (high-mass X-ray 
binaries, or HMXBs, most of which are X-ray pulsars; see Apparao 1994; 
Campana \etal\ 2001), and old, magnetically-accreting white dwarfs 
(polars and intermediate polars; see Ezuka \& Ishida 1999). 

In both classes of source, the compact object has a strong magnetic field
that channels accreted material onto polar caps. If the field is mis-aligned
with the spin axis of the compact object, then the caps can produce 
modulated X-ray emission. The amplitudes of the pulsed emission from
pulsars and polars can be more than 40\% rms
\citep[e.g.,][]{hab98,rr99,nw89}. Therefore, in order
to explore the natures of the hard X-ray sources at the Galactic center, 
we have searched for periodic X-ray variability during a 487-ks exposure
that was taken over the course of two weeks in 2002 May to June. 
In this paper, we report the detection of eight sources that exhibit 
periodic X-ray variability in the direction of the Galactic center. 
We discuss their 
natures based upon the lengths of their periods and the spectral
properties of their X-ray emission. We also report an upper limit on the
optical magnitude of one X-ray source with periodic variability that lies
in the foreground of the Galactic center.

\section{Observations and Data Analysis\label{sec:obs}}

Twelve separate pointings toward the Galactic center have been carried out 
using the Advanced CCD Imaging Spectrometer imaging array 
\citep[ACIS-I;][]{gar02} aboard 
the {\it Chandra X-ray Observatory}, 
in order to monitor \sgrastar\ (Table~\ref{tab:obs}).
The ACIS-I is a set of four 1024-by-1024 pixel CCDs, covering
a field of view of 17\arcmin\ by 17\arcmin. When placed on-axis at the focal
plane of the grazing-incidence X-ray mirrors, the imaging resolution 
is determined primarily by the pixel size of the CCDs, 0\farcs492. 
The CCDs also 
measure the energies of incident photons, with a resolution of 50--300 eV 
(depending on photon energy and distance from the read-out node) within a 
calibrated energy band of 0.5--8~keV. The CCD frames are read out every 3.2~s,
which provides the nominal time resolution of the data.

The methods we used to create a combined image of the field,
to identify point sources, and to compute the photometry for each source
are described in \citet{mun03}. In brief, we created an event list for 
each observation in which we corrected the pulse heights of each event for the 
position-dependent charge-transfer inefficiency \citep{tow00}. We excluded 
non-X-ray background events, events that did not pass the standard 
ASCA grade filters and CXC good-time filters, and intervals of 
strong background flaring. We used the CIAO tool 
\program{acis\_bary} to correct the arrival time of each event to the 
Solar System barycenter. The final total live time was 626 ks. 
We then applied a correction to the 
absolute astrometry of each pointing using three Tycho sources detected 
strongly in each \chandra\ observation \citep[see][]{bag03},
and re-projected the sky coordinates of each event to the tangent plane at the 
radio position of \sgrastar\ in order to produce a single composite image.
The image (excluding the first part of ObsID 1561, during which a 
$10^{-10}$~\ergcms\ transient was observed) was searched for point 
sources using \program{wavdetect} in three energy bands (0.5--8~keV, 
0.5--1.5~keV, and 4--8~keV) using a significance threshold of $10^{-7}$.
We detected a total of 2357 X-ray point sources, of which 1792 were detected 
in the full band, 281 in the soft band (124 exclusively in the soft band), 
and 1832 in the hard band (441 exclusively in the hard band). 

We computed photometry for each source in the 0.5--8.0~keV band using 
the \program{acis\_extract} routine \citep{bro02} from the Tools for 
ACIS Real-Time Analysis 
(TARA).\footnote{\html{www.astro.psu.edu/xray/docs/TARA/}}
We extracted event lists for each source for each observation, using a 
polygonal region generally chosen to match the contour of 90\% encircled energy
from the PSF, although smaller regions were used if two sources were nearby
in the field. We used a PSF at the fiducial energy of 1.5~keV
for foreground sources, while we used a larger extraction area corresponding 
to an energy of 4.5 keV for Galactic center sources. A background event list 
was extracted for each source from a circular region 
centered on the point source, excluding from the event list ({\it i}) 
counts in circles circumscribing 
the 95\% contour of the PSF around any point sources and ({\it ii}) bright, 
filamentary structures. The size of each background
region was chosen such that it contained $\approx 1200$ total events 
for the 12 observations. 
The net counts in each energy band were computed from the 
total counts in the source region minus the estimated background. The
photometry for the complete sample of sources is listed in the electronic
version of Table~3 from \citet{mun03}.


\subsection{Periodicity Search}

We searched for periodicities using data from the 487 ks of observations 
that took place over 14 days between 2002 May 22 and June 4 
(Table~\ref{tab:obs}). These observations are long enough to provide 
good signal-to-noise, and they occur over a short enough amount of time that 
a single ephemeris is likely to be applicable to a signal from any given 
source (although see Section~3).
We searched all 285 sources with more than 100 net counts during
this time interval. 
We did not search for periodicities in the remaining 139 ks of 
data, which were taken in 11--41~ks exposures separated by up to a year. 
It is not possible to identify periods in the individual
observations because the sources are too faint, and combining all of the 
observations would require a prohibitively large search through the parameter 
space of possible ephemerides.

We produced a periodogram of each source using the $Z_1^2$ or Rayleigh 
statistic \citep{buc83},
\begin{equation}
Z_1^2 = {{2}\over{N}} \sum_j \left(\cos^2\phi(t_j) + 
\sin^2\phi(t_j)\right),
\end{equation}
where $N$ is the total number of events, $t_j$ is the arrival time of 
an event, and $\phi$ is the phase at time $t$ that would be expected
for a constant-frequency modulation $\phi(t) = 2\pi\nu t$.
 The lowest frequency ($\nu$) considered was 
$1\times10^{-5}$ Hz, which is approximately the inverse of the length of 
the longest observation. The highest frequency was 0.1 Hz, as the 
photon arrival times are only recorded with an accuracy of 3.2 s.
The frequency resolution was $1\times10^{-6}$ Hz, which is the inverse of 
the total time, $10^6$~s spanned by the observations used to search for 
periodicities. This is the highest resolution with which the periodogram 
can be 
sampled using independent frequency bins. 
For a signal resulting purely from Poisson noise (i.e., a white-noise 
signal), $Z_1^2$ has a 
chi-squared distribution
with two degrees of freedom. This statistic is useful in a case like 
ours when the 
arrival times of individual photons are well-known, and yet there are too
few photons to analyze evenly-binned data. 
This method also naturally ignores gaps in the data produced when the 
Galactic center was not observed. We performed Monte-Carlo simulations that
indicate that 
these gaps do not cause significant noise leakage due to the sampling 
window function (see Press \etal\ 1992 pp. 437--447 for a general 
discussion of noise leakage). 
Red noise due to long term variability is apparent below 
$5\times10^{-5}$~Hz in the power spectra of $\approx 10$\% of the sources, 
but does not otherwise affect the detectability of coherent signals at higher
frequencies. 

Each of the 285 sources was searched for periodicities in 
approximately $\approx 10^5$ frequency bins, so $3\times10^7$ trial 
frequencies were examined in total. Therefore, there is only a 
1\% chance that Poisson noise would produce a signal stronger than 
$Z_1^2 > 43.5$ in our entire search. 
We consider any signal stronger than this to be a secure detection of a 
periodicity.
The rms amplitude $A$ of a sinusoidal signal can be computed from 
$Z^2_1$ as 
\begin{equation}
A = \left({{Z^2_1} \over {N}} \right)^{1/2} {{N} \over {N-B}},
\label{eq:rms}
\end{equation}
where $N$ is the total number of events from the extraction region, and
$B$ is the number of background counts \citep[e.g.,][]{lea83}.
Given the typical background of 30
\begin{inlinefigure}
\centerline{\epsfig{file=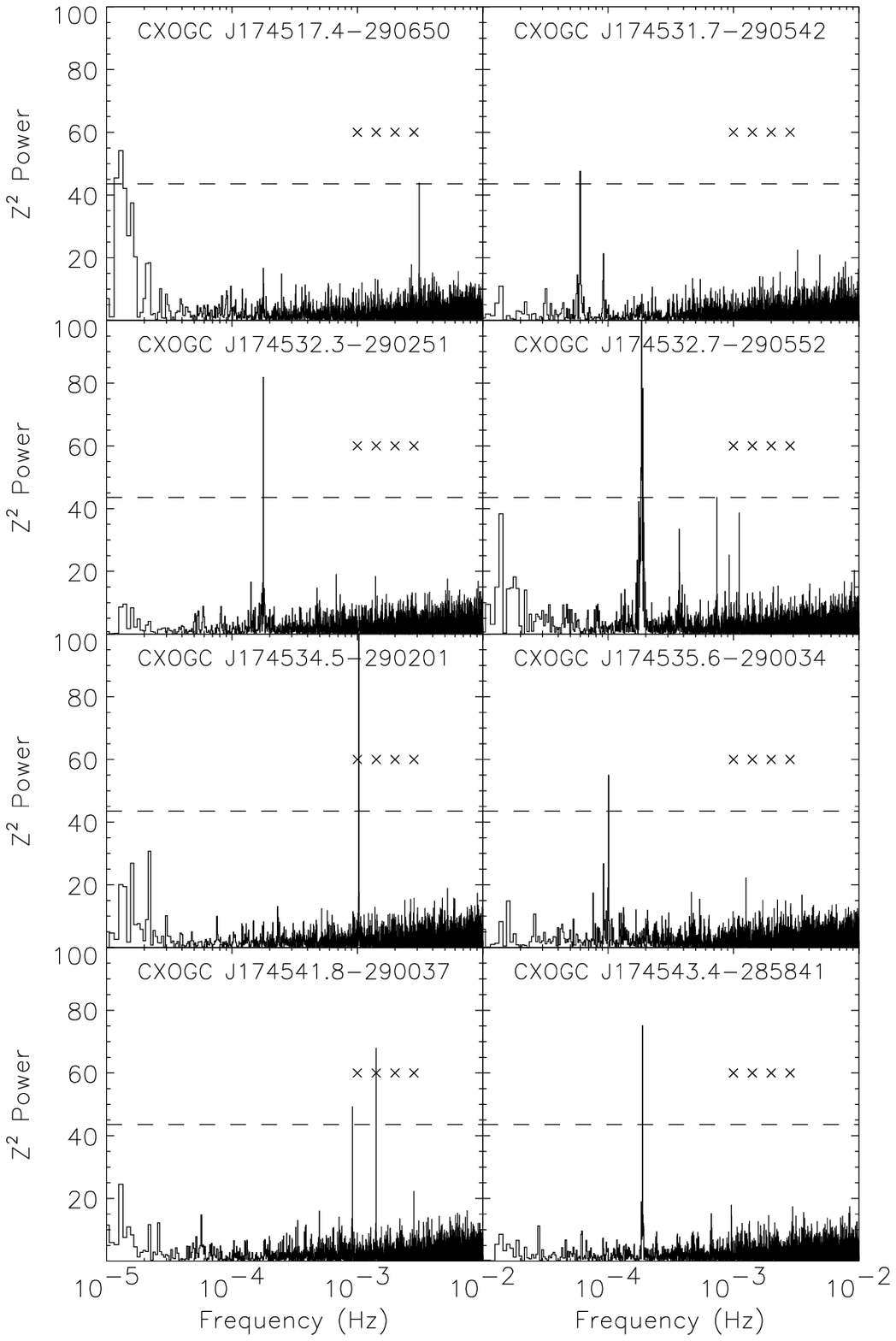,width=0.95\linewidth}}
\caption{
Periodograms of the eight sources in which significant periodicities are
detected, computed using the $Z^2_1$ statistic. We have only plotted 
frequencies between $10^{-5}$ and $10^{-2}$~Hz, since between $10^{-2}$ and
$10^{-1}$ the power is dominated by noise. The 
dashed lines indicate the 99\% confidence level for detecting a signal 
given the total number of sources and frequency bins searched. 
The $\times$'s indicate
frequencies at which spurious signals from the satellite dither would be 
expected
if the source were near a chip gap or bad column. CXOGC~J174541.8--290037
exhibits a signal at the pitch frequency of $1.415\times10^{-3}$~Hz, in 
addition to the intrinsic signal at $9.16\times10^{-4}$~Hz.
}
\label{fig:per}
\end{inlinefigure}

\noindent
counts for a 
source within a few arcminutes of the aim-point, we would be able to 
detect a completely modulated signal (70\% rms) with only 100 counts 
from a source.

We detected 23 significant periodicities. We then checked to ensure that 
none of them were at multiples of the frequencies at which the pointing 
of the satellite is dithered. The dither is designed to distribute photons 
over many CCD pixels, and has a period of 706.96~s in pitch 
and 999.96~s in yaw. The count rate from a source that lies near 
a chip gap 
or bad column would appear to vary with one or both of the above periods.
There were 16 sources that exhibited variations at multiples of the dither 
frequencies, all of which lay near chip gaps or bad columns. Significant 
periodicities that were clearly not associated with the dither frequencies
were detected from 8 sources (including one that also exhibited variations
at a dither frequency). The locations of these sources on the ACIS-I CCDs 
did not suggest any instrumental anomalies
\begin{inlinefigure}
\centerline{\epsfig{file=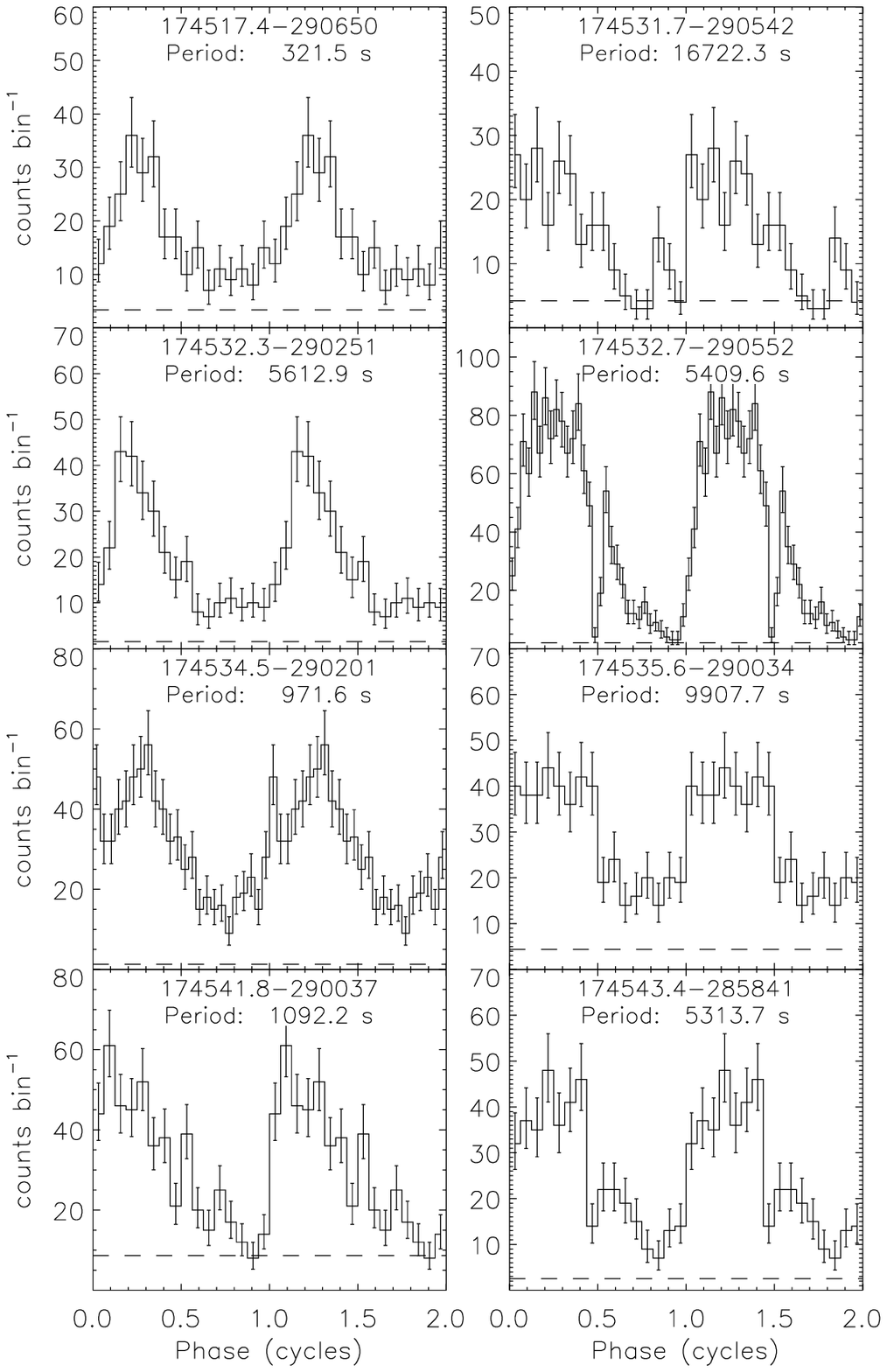,width=0.95\linewidth}}
\caption{Folded profiles of the eight sources with significant X-ray 
periodicity, repeated for two cycles for clarity. The background 
level is indicated with a dashed line. CXOGC~J174517.4--290650,
CXOGC~J174532.3--290251, and \forbit\ exhibit significant harmonic content,
while the profiles of the remaining sources are consistent with sinusoids.
The uncertainties are 1 $\sigma$ upper and lower bounds defined according to 
\citet{geh86}.
}
\label{fig:fold}
\end{inlinefigure}

\noindent
that might produce periodicities, so 
we concluded that the periods are intrinsic to the sources themselves. 
We listed these eight sources in Table~\ref{tab:per}, and displayed their 
power spectra in Figure~\ref{fig:per}. 

We found possible periodicities in three more 
sources, but these require further confirmation to be considered secure. Two 
sources appear to produce highly significant coherent signals at very low 
frequencies: CXOGC~J174540.1--290055 ($1.5 \times 10^{-5}$~Hz), and
CXOGC~J174552.2--290744 ($1.4 \times 10^{-5}$~Hz). Although the power 
appears to be strongly peaked at the above frequencies, and there is little
power elsewhere below $5 \times 10^{-5}$~Hz, it cannot be ruled out that 
these periodicities result from red noise. Moreover, the periods are so long 
that fewer than 3 cycles could be detected in the longest individual 
observation. The third source, CXOGC~J174546.2--285906, exhibits a 
coherent signal at $6 \times 10^{-5}$~Hz with $Z_1^2 = 37.7$. This signal
has an 18\% chance of occurring due to Poisson noise given our entire search, 
so we do not include it with the secure detections in Table~\ref{tab:per}.
\begin{figure*}[thb]
\centerline{\epsfig{file=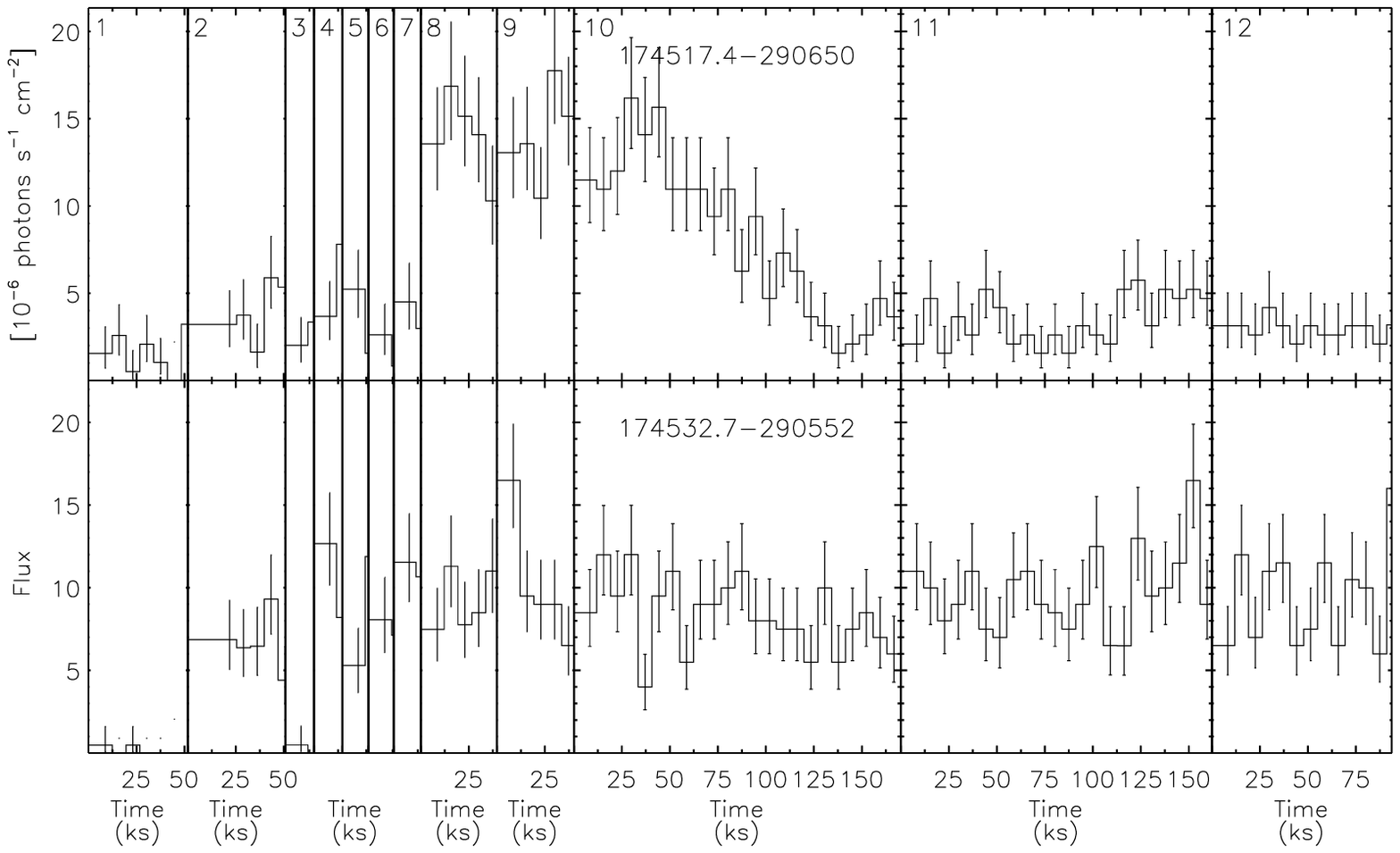,width=0.9\linewidth}}
\caption{ Photon fluxes from the two sources that exhibit 
significant long-term variability, computed by dividing the count rate 
by the mean value of the ARF and by the exposure in each bin.
The background level has not been subtracted. Data from the individual 
observations are 
plotted in 7200-s bins in order to illustrate the short-term variability; 
the number of each observation from Table~1 is indicated at the top. 
The total time over which the data were taken spans nearly three years. 
The flare from CXOGC~J174517.4--290650 was visible for three weeks.
The energy flux can be estimated by assuming the spectrum is a $\Gamma = 0.5$ 
power law absorbed by a column of $6\times10^{22}$ cm$^{-2}$, for which 
the mean energy per photon is $8\times10^{-9}$~erg (2--8~keV).
The uncertainties are 1 $\sigma$ upper and lower bounds defined according to 
\citet{geh86}.
}
\label{fig:lcurve}
\end{figure*}
Unfortunately, the remaining observations (1--7 in Table~\ref{tab:obs}) did 
not provide enough counts to confirm these possible signals.

In order to determine the period of the variability for each source
in Table~\ref{fig:per} more accurately, we 
computed
 $Z^2_1$ over a frequency range of $2\times10^{-6}$~Hz around 
the peak power in the periodogram using a frequency resolution of
$10^{-7}$ Hz. This over-samples the frequency 
resolution for which each measurement of the power is independent by a 
factor of 10, allowing us to precisely determine the frequency at which the
largest power is produced. The corresponding periods are listed in 
Table~\ref{tab:per}. We then produced 
light-curves folded at the period obtained from the peak of the over-sampled
periodogram, which are displayed in Figure~\ref{fig:fold}.
We measured the amplitudes and ephemerides of the periodicities by fitting a
sine function to the folded profile, and listed these quantities in 
Table~\ref{tab:per}.
Therefore, the amplitude is that of the fundamental sinusoidal component
of the variation, which is useful for comparison to amplitudes and 
upper limits detected in periodicity searches. The ephemeris is defined 
by the zero phase of the sine function closest to the start of the observation,
and is reported in Modified Julian Days, Barycentric Dynamical Time.
We determined the 1 $\sigma$ uncertainties on the parameters listed in 
Table~\ref{tab:per} using a Monte-Carlo simulation, in which we produced 
100 light curves that matched the measured count 
rate, sampling window, and period and amplitude of the sinusoidal signal
for the data 
from each source, and analyzed them in the same manner as the real events.

We next characterized the shapes of the modulations by computing
Fourier power spectra of the folded profiles from each source in 
Figure~\ref{fig:fold}. Power at the first harmonic of the main signal
was present in CXOGC~J174517.4$-$290650 and CXOGC~J174532.3$-$290251
with a single-trial probability of 0.7\% that they were due to noise 
(power $p = 10$ using the normalization of Leahy \etal\ 1983), and from 
CXOGC~J174532.7$-$290552 with a single-trial chance probability of $10^{-11}$
($p = 50$). The harmonic signals from the other sources had a single-trial
probability of at least 5\% that they were due to noise. We converted the 
powers of the 
harmonic signals to fractional rms amplitudes using Equation~\ref{eq:rms},
substituting the Fourier power for 
$Z_1^2$ \citep{lea83}. We also computed
$1 \sigma$ uncertainties on the amplitudes, taking into account the 
expected distribution of power from Poisson noise \citep{vau94}. 
In all three cases, the first harmonics had a rms amplitudes of 
$20\% \pm 5\%$. We found no evidence for higher harmonics. We also folded the 
data about $1/2$ the period of the strongest signal in order to search for 
evidence that the main signals are produced by two magnetic
poles on the compact object. Marginally significant power was detected 
at half the period of the main signal only from CXOGC~J174535.6$-$290034,
with a 2\% chance that it was due to Poisson noise for a single trial 
($p = 7$).  

We also searched for energy dependence in the amplitudes and shapes of the 
pulse profiles. We derived folded profiles for each source in the 
0.5--4.7~keV and 4.7--8.0~keV bands, and computed Fourier power spectra of
the resulting profiles to quantify their amplitudes and harmonic content. 
We find that the pulse profiles from three sources are inconsistent with a 
constant amplitude at the $2.5 \sigma$ level. The modulation 
amplitudes decrease with energy in CXOGC~J174517.4$-$290650 from 
$(74 \pm 11)\%$ to $(36\% \pm 9)\%$, and in CXOGC~J174543.4$-$285841
from $(62 \pm 9)\%$ to $(34 \pm 6)\%$ (uncertainties are $1 \sigma$). 
The modulation amplitude increases 
with energy in CXOGC~J174541.8$-$290037 from $(29 \pm 7)\%$ to
$(53 \pm 6)\%$. Given that the profiles of these three systems contain only 
100--200 net counts in each energy band, it would be useful to confirm these 
results with future observations. The amplitudes of the variations for the 
other sources were constant with energy to within their $1\sigma$ 
uncertainties. We find no evidence for changes in the shapes of 
the profiles as a function of energy.

Finally, in order to search for variability in each source aside from the 
periodic signals, we extracted light curves using the CIAO tool
\program{lightcurve}. Two sources exhibited long-term X-ray variability,
so we display them in Figure~\ref{fig:lcurve}. One of the two,
\forbit, appears fainter in the observations in 1999 September (ObsID 0242)
and 2001 July (ObsID 1561, part 2) than 
in the remaining observations, but its mean flux was constant during 
the observations in 2002 May and June.
The other, CXOGC~J174517.4$-$290650, decreased from a flux of 
$9\times10^{-14}$~\ergcms\ to $8\times10^{-15}$~\ergcms\ 
during ObsID 3392 (we have assumed 
a $\Gamma = 0.5$ power law absorbed and scattered by 
$N_{\rm H} = 6\times10^{22}$ cm$^{-2}$ of gas and dust). 
We searched for pulsations from this source separately 
in the first two and last two observations, when the mean flux from the 
source did not vary strongly. The pulsations are detected 
strongly in ObsIDs 2943 and 3663, and their properties are listed in 
Table~\ref{tab:per}. The oscillations are not detectable in ObsIDs 3393 and 
3665 due to the low count rate; with only 77 net events from the source
against 140 background events, a fully modulated sinusoidal signal would be 
undetectable (Equation~\ref{eq:rms}).

\subsection{Spectra}

In order to compare the Galactic center sources with periodic X-ray variability
to known classes of objects, we extracted their phase-averaged energy spectra 
from the entire 626 ks data set.
We computed the effective area function (ARF) at the position of each 
source for each observation. This was corrected to account for the fraction 
of the PSF enclosed by the extraction region and for the variable hydrocarbon 
build-up on the 
detectors.\footnote{\html{cxc.harvard.edu/cal/Acis/Cal\_prods/qeDeg/}}
We used response matrices appropriate for data corrected to account for 
charge-transfer inefficiency. We used the average of the effective area and
response functions, weighted by the number of counts from each source in 
each observation.

We modeled the X-ray spectra in \program{XSPEC}, using a power-law
continuum absorbed at low energies by gas (using the model {\tt phabs} in 
\program{XSPEC}) and dust (using a modified version of the model 
{\tt dust} in which the approximation that the dust was optically thin 
was removed). The column depth of dust was set to 
$\tau = 0.485 \cdot N_{\rm H}/(10^{22} {\rm cm}^{-2})$ \citep{bag03}, 
and the halo size to 100 times the PSF size.
The spectra with the resulting fits are displayed in 
Figure~\ref{fig:spec}. Half of the sources could not be adequately fit with 
this simple model; from Figure~\ref{fig:spec} it is clear that this is 
at least partly due to excess counts between 6--7~keV, where one expects
emission from both neutral and 
highly-ionized iron. Therefore, we  
added Gaussian lines 
\begin{inlinefigure}
\centerline{\epsfig{file=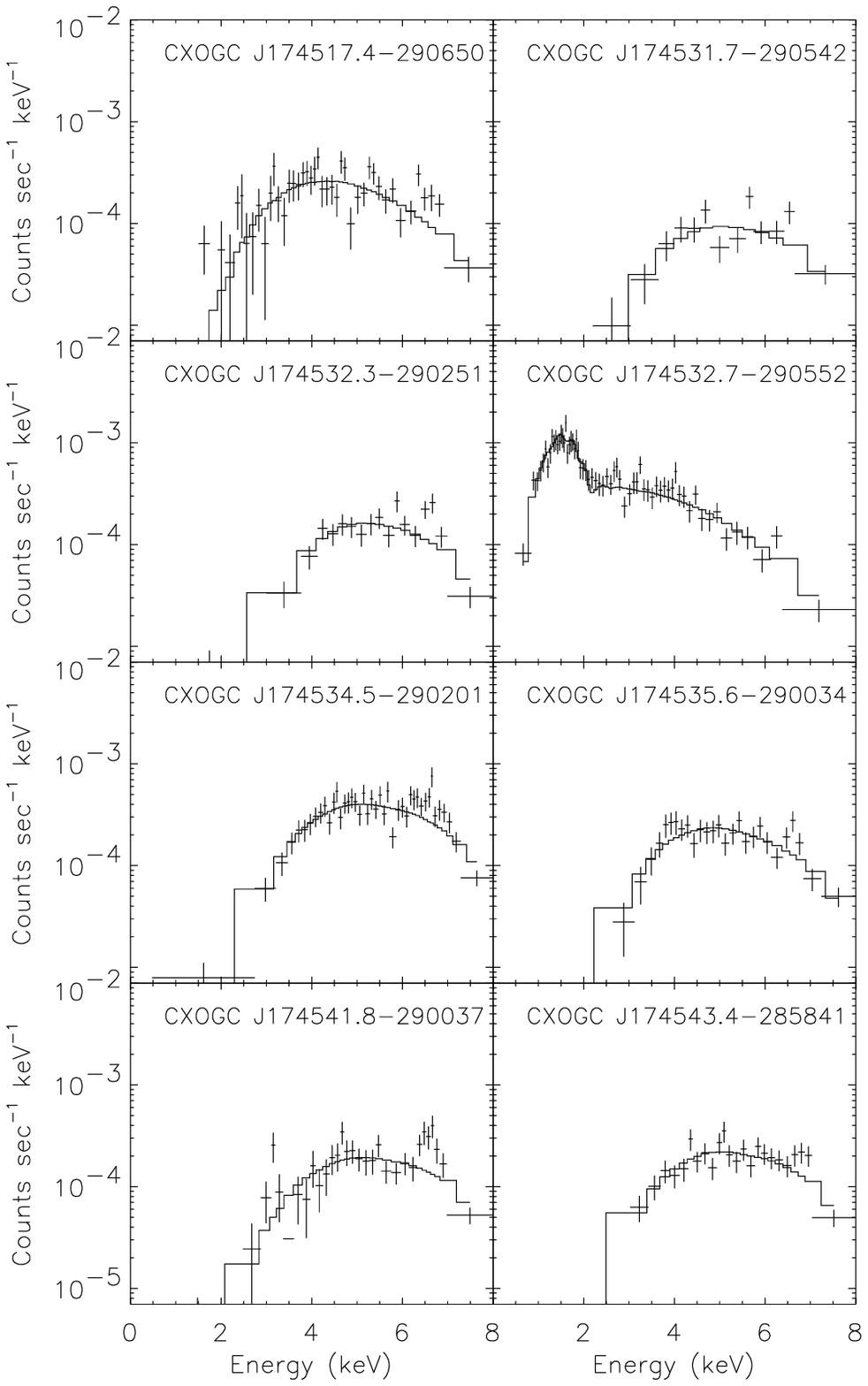,width=0.95\linewidth}}
\caption{
X-ray spectra of the eight sources with periodic X-ray variability.
Each spectrum is displayed in units of detector counts s$^{-1}$ 
keV$^{-1}$ as a function of energy in keV, so that the varying 
effective area of the detector is convolved with the spectrum. 
The solid histograms represent the best-fit absorbed power laws. 
The power laws are statistically acceptable in four cases, while
in the remaining four there are significant residuals near 6.7~keV, where
emission from highly ionized iron is expected (see Table~\ref{tab:spec}).
\forbit\ is in the foreground of the Galactic center, as evident from the
copious emission below 2~keV. The remaining sources lie near or beyond the 
Galactic center, and exhibit hard ($\Gamma<1$) power laws.
}
\label{fig:spec}
\end{inlinefigure}

\noindent
to the models for these four sources, which significantly
improved the fits. The resulting spectral parameters are listed in 
Table~\ref{tab:spec}. 

The column of gas and dust toward all but one of the sources is comparable to
or greater than the value toward \sgrastar, 
$6\times10^{22}$ cm $^{-2}$ \citep{bag03}. The remaining source, 
\forbit, is
clearly located in the foreground of the Galactic center. This source
also has the softest spectrum, which is consistent with either a 
$\Gamma = 1.4$ power law or a $kT = 25$~keV bremsstrahlung continuum.
The remaining sources either have hard X-ray
spectra that are consistent with $\Gamma < 1$ power laws, or have 
poorly constrained $\Gamma \sim 0-2$ continuum spectra with evidence for 
Fe emission lines. 
If we use a bremsstrahlung 
continuum spectrum instead, we 
only obtain lower limits on their temperatures of about 30~keV.
We note that this does not necessarily imply that the spectra are 
intrinsically that flat. Given that the absorbing columns toward several 
of the sources are larger than the Galactic value, it
is possible that absorbing material local to each of these sources 
partially covers the X-ray emitting regions \citep[compare][]{app94,ei99}. 
This can make a spectrum look much 
flatter than it is intrinsically. Such a ``partial-covering
absorber'' model is often used for magnetic CVs, but we feel that using this 
more complicated model for the current data is unwarranted given the 
small number of counts recorded for each source.

Emission lines from Fe are present in at least four of the Galactic center 
sources with periodic variability (Table~\ref{tab:spec}). 
The equivalent widths of the lines range from 600--1000~eV for the sources 
for which an absorbed power law alone cannot adequately model the spectrum. 
The lines are all located at 
approximately 6.7~keV, indicating that they result from He-like Fe. Likewise,
if we add a Gaussian line at 
$\approx 6.7$~keV to those Galactic center sources for which an absorbed 
power law does provide an adequate fit, we find that their best-fit 
equivalent widths are 400--700~eV. Although the equivalent widths of 
these lines are similar to those observed
from the diffuse X-ray emission from the Galactic center (Park \etal, in
preparation), the construction of the background regions ensures that the 
line emission originates from the point sources. We have also 
confirmed that each of the sources with strong 6.7~keV emission is visible
as a point source in an image produced from 6.5--7.0~keV X-rays. 
On the other hand, any ionized Fe emission from the 
foreground source \forbit\ must be weak, as the 90\% upper limit on the 
equivalent width of a line at 6.7~keV is 100~eV.

The lines detected from the Galactic center sources were resolved with widths 
of 0.14--0.3~keV in three out of four cases. Several factors could 
contribute to the width of the lines: they could be a blend of several 
ionization species, Doppler-broadened by the motions of the accreted material, 
or broadened by Compton scattering in the accretion flow
(e.g., Hellier, Mukai, \& Osborne 1998). 

\section{Discussion}

We have searched for periodicities in 285 sources detected with more than 
100 net counts during 487~ks of \chandra\ observations of the Galactic 
center taken from 2002 May 22 to June 4. In eight of these sources, we have 
found significant variability with periods ranging from 320~s to 
4.5 h and amplitudes ranging from 40--70\% rms. 
Our search for pulsations was motivated by the prevalence of sources
with hard ($\Gamma < 1$) power-law spectra in the \chandra\ image
of the Galactic center, which we have tentatively identified as 
either magnetically-accreting white dwarfs or wind-accreting neutron stars 
(Muno \etal\ 2003; see also Apparao 1994; Ezuka \& Ishida 1999).
We have illustrated which of the sources are observed
with periodicities in Figure~\ref{fig:cint}, in which we plot the net 
number of counts (0.5--8.0~keV) from each source versus a hard color that 
parameterizes the steepness of the spectrum. The
hard color is defined as $(h-l)/(h+l)$, where $h$ is the number
of counts from 4.7--8.0~keV and $l$ is the number of counts
from 3.3--4.7~keV. Not surprisingly, the brightest 
sources are more likely to be detected with periodicities, since the minimum 
detectable pulsed fraction decreases as the net number of counts from a 
source increases (Equation~\ref{eq:rms}). Moreover, seven of
the sources with periodic variability have hard colors $> 0$
(Fig.~\ref{fig:cint}), which are consistent with $\Gamma < 1$ power-law 
spectra (Table~\ref{tab:spec} and Figure~\ref{fig:spec}). This supports
the suggestion that many of the hard sources are either magnetic CVs or 
HMXB pulsars. These seven sources also are heavily absorbed, which suggests
that they are located at or beyond the Galactic center (Table~\ref{tab:spec}).
The eighth source is a relatively soft foreground source, 
and will be discussed in detail in Section~3.2. 

\subsection{The Natures of the Galactic Center Sources}

It is difficult to determine whether the Galactic center 
systems with periodicities are magnetic CVs or HMXB pulsars based
on the X-ray data alone. The luminosities of the Galactic center sources
range from $(1 - 5) \times 10^{32}$~\ergsec
(2--8~keV at 8 kpc). 
These luminosities are comparable to those of the brightest observed magnetic 
CVs \citep{ver97,ei99,gri01,pool02}, but are a factor of 
$\approx$10 
below the luminosities of the faintest known HMXB pulsars \citep{neg00,cam01}.
However, HMXBs are less common in the Galaxy than magnetic CVs 
(compare Pfahl, Rappaport, \& Podsiadlowski 2002; Howell \etal\ 2001), and
most HMXBs are identified through transient outbursts that reach luminosities
of $10^{38}$~\ergsec. Therefore, the known HMXBs lie at much greater 
distances than the known CVs --- the typical 
distance for an HMXB is $\sim 8$~kpc, compared to $\sim 100$ pc for a 
magnetic CV \citep{app94,war95}.
The lack of known HMXBs with low luminosities may be a selection effect,
and the hard Galactic center sources could represent the first such systems
identified \citep{prr02}.

The spectra of magnetic CVs and HMXB pulsars also can appear 
similar between 2--8~keV. Both types of systems can often be described 
with a $\Gamma < 1$ power-law continuum, and both exhibit iron emission 
between 6--7~keV with equivalent widths up to $\sim 500$~eV 
\citep{app94,hmo98,ei99}. The Fe 
emission from three of the Galactic center sources also appears to be resolved
(Table~\ref{tab:spec}).  Likewise, Fe emission was
resolved with widths of $\sim 200$~eV from about half of the magnetic CVs 
studied using the \asca\ SIS by \citet{hmo98}.
To our knowledge, 6.7~keV Fe lines this broad have not been reported from 
HMXB pulsars, although this may be because few comparable studies 
of the Fe lines from such systems have been carried out using CCD-quality 
spectra.

The properties of the modulations observed from the Galactic center
sources suggest that they originate from the 
spin periods of either magnetically-accreting white dwarfs or
wind-accreting neutron stars.
The modulation amplitudes in Table~\ref{tab:per} range from 40\% to 70\% rms.
Similarly, a survey of the literature yields a range of 5--40\% 
rms for the amplitudes of pulsations from magnetic CVs 
\citep[e.g.,][]{nw89,sch02,rc03}, and 5--50\% rms for HMXB pulsars
\citep{rap82,hab98,rr99,isr00,hal00}. We have also detected significant 
non-sinusoidal components 
\begin{figure*}[thb]
\centerline{\epsfig{file=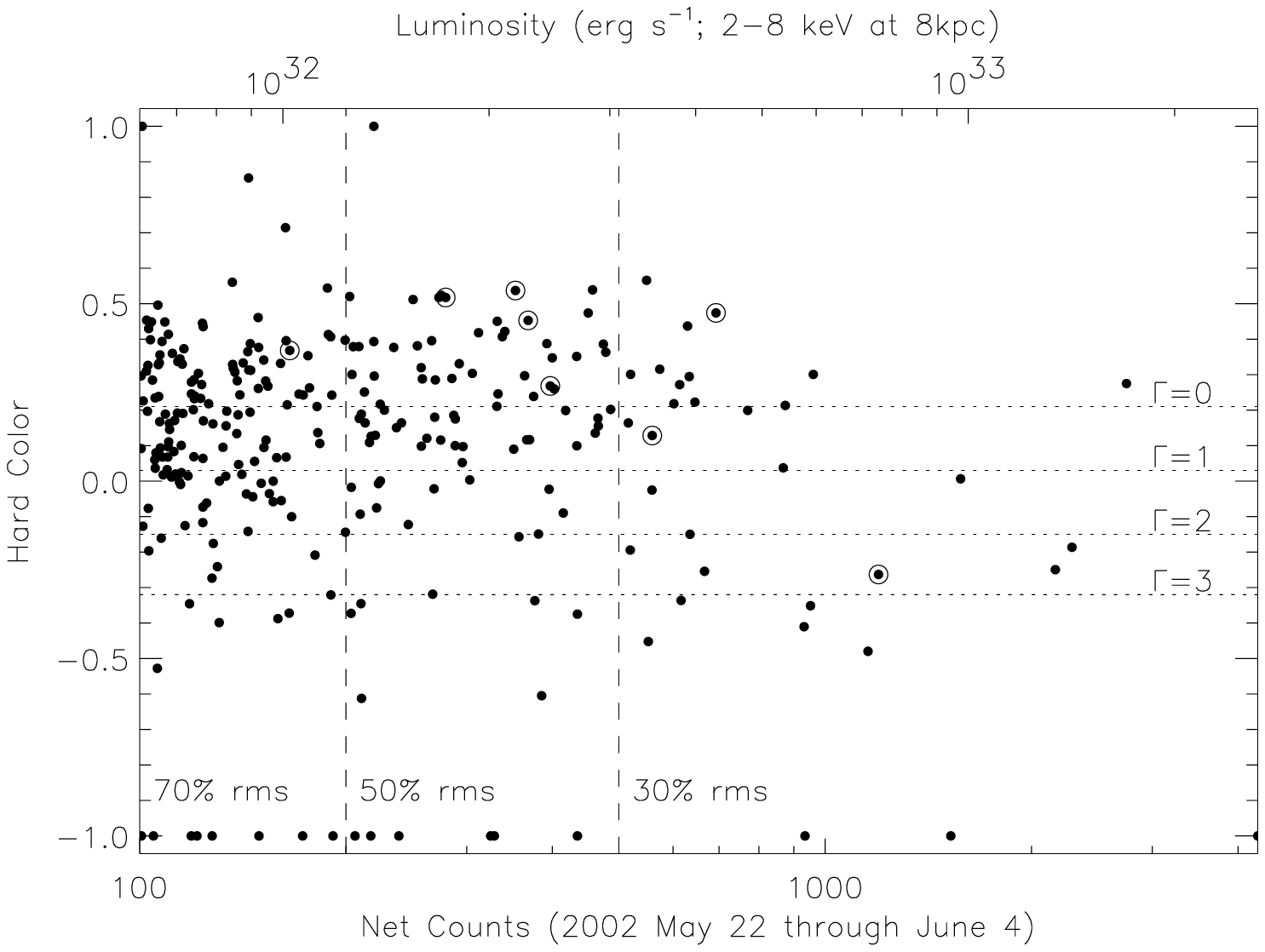,width=0.75\linewidth}}
\caption{
Hard colors of the 285 sources searched for periodicities plotted as a
function of the net counts (0.5--8.0~keV) from each source during the 7
observations taken between 2002 May 22 and June 03. The
hard color is defined as $(h-l)/(h+l)$, where $h$ is the number
of counts from 4.7--8.0~keV and $l$ is the number of counts
from 3.3--4.7~keV. The luminosities of the sources 
are indicated by the axis at the top of the figure, where we have 
assumed $1~{\rm count}/487~{\rm ks} = 8 \times 10^{-17}$~\ergcms
and a distance of 8 kpc. The conversion was computed 
using \program{PIMMS} for a 
$\Gamma = 0.5$ power-law spectrum absorbed by $6 \times 10^{22}$ cm$^{-2}$
of gas and dust. It is only approximate 
because it does not take into account vignetting or the dither in the 
satellite pointing. 
 The vertical 
dashed lines indicate the numbers of net counts needed 
to detect a modulation with fractional rms amplitude of 50\% and 30\% on top
of 30 counts of background;
a fully modulated signal (70\% rms) can just be detected from the sources 
with 100 net counts in this diagram.
The horizontal dotted lines indicate the hard colors corresponding to 
power-law spectra over a range of photon index $\Gamma$, assuming an 
absorption column of $6\times 10^{22}$ cm$^{-2}$.
Sources above the $\Gamma =1$ line are likely to be magnetized CVs or HMXB 
pulsars located at the Galactic center. 
Sources with detected periodicities are marked additionally with open 
circles, in 
order to convey the fraction of hard sources that are detected with 
oscillations as well as the number of sources in which oscillations of
a given amplitude could have been detected.
}
\label{fig:cint}
\end{figure*}
in the profiles of three sources in 
Figure~\ref{fig:fold} (see Section~2.1), which is commonly
seen from both HMXB pulsars and magnetic CVs \citep[e.g.,][]{rap82, rc03}. 
Unfortunately, aside from in \forbit\ (see Section 3.1),
there is not enough signal from the Galactic center sources to speculate in 
detail on how the pulse shape is formed.

The periods of the modulations 
are also consistent with the spin periods of white dwarfs in CVs and
neutron stars in HMXBs. The median spin period of the 112
magnetic CVs in the catalog of \citet{rk03} is $P \approx 6000$~s. 
All of the periods listed in Table~\ref{tab:per} are comfortably in the 
range of those observed from magnetic CVs, although only 3 such systems 
are known with periods shorter than that of CXOGC~J174517.4$-$290650,
$P = 321$~s. In contrast, the 68 Galactic HMXB pulsars in the catalog of 
\citet{liu00} have a median $P \approx 91$~s, and only 3 have $P > 900$~s. 
Although this suggests that the sources detected with periodicities at the 
Galactic center are more likely to be magnetic CVs, it should be emphasized 
that there is 
no lower limit to how slowly a wind-accreting neutron star can rotate. 
Indeed, the slowest pulsar known, 4U~0114+650, has a period of over 
$10^4$ s \citep{hal00}, which is comparable to the longest periods listed 
in Table~\ref{tab:per}. 

Finally, our ability to use the seven sources with periodicities to 
draw conclusions about the entire sample of hard sources at the Galactic 
center is limited, because the Doppler motion of a neutron star in its binary 
orbit can prevent us from detecting the short-period modulations that are
typical of HMXB pulsars. This Doppler motion would smear the 
power in a coherent pulsar signal over several adjacent frequency bins. 
For a circular orbit, the shift in frequency is approximately
\begin{equation}
{{\Delta\nu}\over{\nu}} = 1.5\times10^{-3} 
{{M_{\rm C} \sin i}\over{M_{\rm tot}^{2/3} P_{\rm day}^{1/3}}} 
\left(1-\cos\left[{{2\pi\Delta T}\over{P}}\right]\right),
\end{equation}
where $M_{\rm C}$ is the mass of the companion star in solar masses,
$i$ is the inclination of the binary, $M_{\rm tot}$ is the total mass 
of the system, 
$P_{\rm day}$ is the orbital period in days, and $\Delta T$ is the exposure
time.
For a massive X-ray binary, $M_{\rm C} \approx M_{\rm tot} \approx 10$\msun.
The orbital period can be shorter than 1 day in supergiant HMXBs, but
the more common Be systems tend to have orbital periods between 10--300~days
\citep{liu00}.
For a 10-day orbit, a 10~\msun\ companion, and an inclination of 
60$^\circ$, the power from pulsar signals faster than $\approx 2600$~s 
would be 
spread over two frequency bins. Therefore, binary motion easily could 
explain why we did not detect any signals faster than $10^{-2}$~Hz, which 
would be expected from HMXB pulsars.
%

Thus, the long periods of the seven Galactic center
\begin{figure*}[thb]
\centerline{\epsfig{file=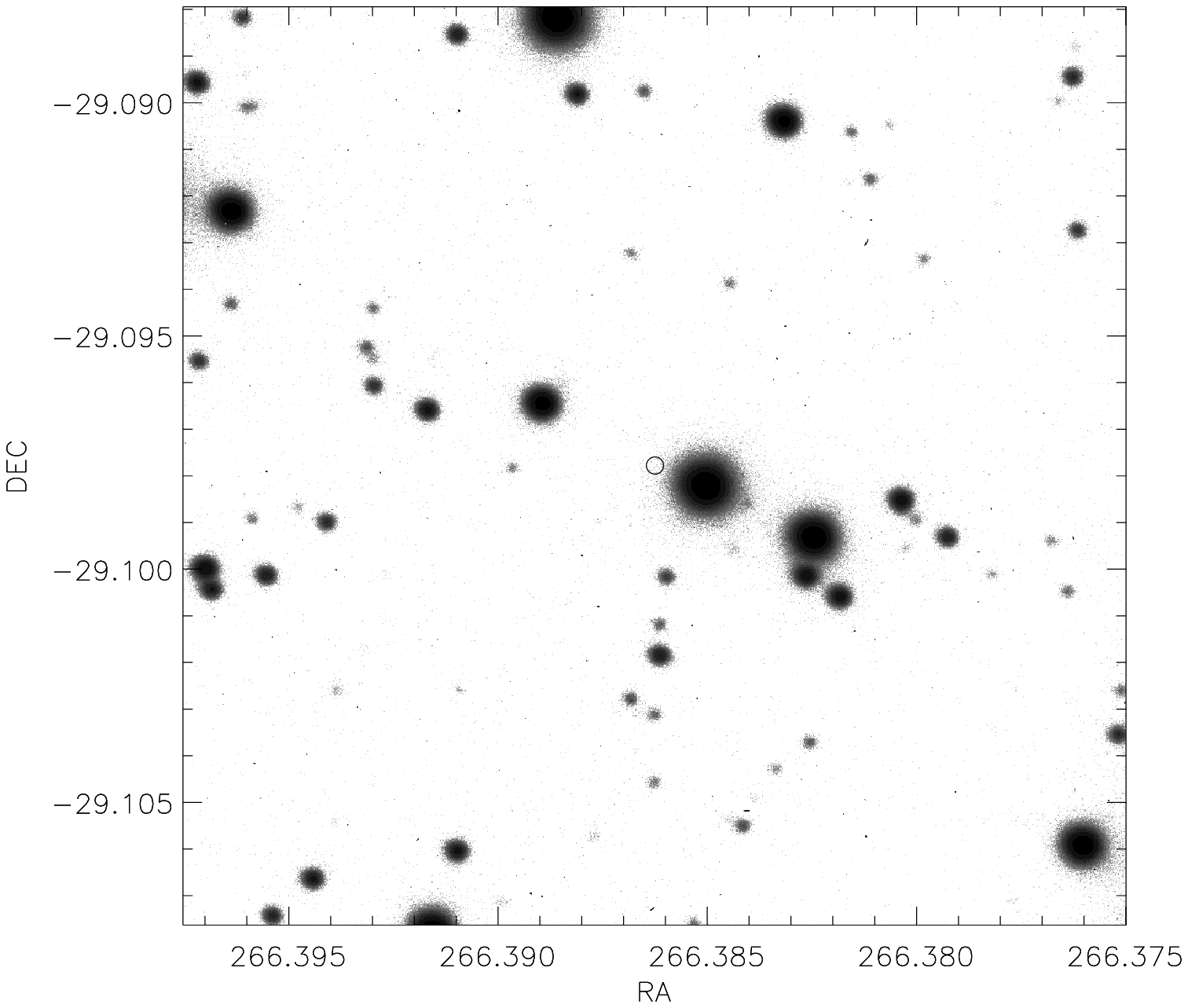,width=0.75\linewidth}}
\caption{
Image of the 2.4 arcmin$^2$ field around \forbit\, taken from a 
5 min $i^\prime$ band exposure using MagIC on the Clay telescope (Magellan-II).
The 1 $\sigma$ error circle for \forbit\ is marked; there is no 
counterpart down to $i^\prime \approx 23$ mag.
}
\label{fig:optical}
\end{figure*}
 sources with 
modulations provide marginal evidence that they are magnetic CVs. However,
the possibility that they are HMXB pulsars cannot be ruled out, and 
the difficulty of detecting short-period modulations from HMXBs
prevents us from extrapolating this conclusion to the 
hundreds of hard sources at the Galactic center for which we have not 
detected significant periodic variations. Indeed, it seems likely that 
other sources with periodic variability below our detection threshold are
present in the image. In Figure~\ref{fig:cint}, 
we have indicated with vertical dotted lines the approximate 
total numbers of counts required to detect sources with 30\% and 50\% rms 
sinusoidal modulations. Only the magnetic CVs and HMXB pulsars with the 
largest-amplitude modulations would have been detectable in our 
search, which suggests that periodicities could be detected from 
many of the hard sources in a deeper exposure.

\subsection{CXOGC~J174532.7--290552 $=$ RXJ 1745.5$-$2905}

The foreground source, \forbit, was first identified as the \rosat\ source 
RXJ 1745.5$-$2905 \citep[source 11 in][]{pt94}, although an X-ray 
periodicity has not previously been reported from the source.
The folded light curve of \forbit\ in Figure~\ref{fig:fold}
exhibits two distinct features
that recur with the same period: a gradual, large-amplitude modulation 
in the total
X-ray flux, and a sharp, 250~s eclipse during which the entire X-ray 
emitting region is obscured. 
These features demonstrate that this source
is a polar (AM Her-type system), in which the spin period of the white 
dwarf is 
synchronized to that of the orbital period. The smooth, large amplitude 
modulation is produced by the rotation of the white dwarf, which causes
the hot region on the white dwarf onto which the accreted material is 
magnetically channeled to be periodically obscured. 
The short eclipse occurs because
we observe the system along the orbital plane with an inclination 
$i > 75^\circ$, so the companion star occults 
the much smaller white dwarf \citep[see][for a review]{war95}. 

%

The absorption column toward the source, $N_{\rm H} \approx 5\times10^{21}$ 
cm$^{-2}$, can be used to estimate its distance roughly. If we assume that 
the density of gas and dust 
in the Galactic disk is distributed radially as an exponential with 
a scale length of 2.7 kpc (Kent, Dame, \& Fazio 1991)\nocite{kdf91} 
and that its total column to the Galactic center distance of 
8~kpc is $6\times10^{22}$ cm$^{-2}$ \citep{bag03}, then we find that \forbit\
lies at a distance of $\approx 2.5$~kpc. This would rank 
\forbit\ with the recently-identified globular cluster CVs 
\citep[e.g.,][]{gri01,pool02,edm03} as one of the most 
distant polars ever identified; the only CVs otherwise identified at such 
large distances are classical novae and 
super-soft X-ray sources \citep[compare][]{war95}. We can infer that its
X-ray luminosity is $\approx 4\times10^{30}$\ergsec, which is well
within the range commonly observed from polars \citep{ei99}. 

Since this source is in the foreground of the Galactic center, we have 
attempted to identify its optical counterpart. An observation of 
the field around \forbit\ was obtained using the 
Raymond and Beverly Sackler Magellan Instant Camera (MagIC) on the 6.5 m 
Clay (Magellan-II) telescope at Las Campa\~{n}as Observatory in Chile.
MagIC is a $2048\times2048$ SITe CCD with
a plate scale of 0\farcs069 pixel$^{-1}$ and a field of view of 
2.36 arcmin$^2$. MagIC was installed at the 
west f/11 Nasmyth focus of the telescope. 
The observation occurred during the first science run on 2002 September 7.
The field was exposed for 5 min behind a Sloan $i^\prime$ filter 
\citep{fuk96}. 
The seeing was 0.8\arcsec (FWHM). The resulting image, corrected 
for bias variations between the 4 CCDs, is displayed in 
Figure~\ref{fig:optical}. Unfortunately, no standard star was observed 
in the $i^\prime$ band because of technical problems later in the evening.

No counterpart was found within 1.5\arcsec\ of \forbit\ 
(Figure~\ref{fig:optical}). Without a 
standard star for calibration, we can only estimate an upper limit of 
$i^\prime < 23$ mag for the intensity of \forbit\ based on exposures taken
with MagIC with similar seeing and duration. The column density of absorbing 
material toward the source, $H_{\rm H} = 5\times10^{21}$ cm$^{-2}$, 
implies a visual 
reddening of $A_V = 2.8$ magnitudes \citep{ps95}, which translates to an 
extinction
near the $i^\prime$ band of $A_I = 1.3$ magnitudes \citep{rl85}. Thus, 
the deabsorbed $i^\prime$ magnitude 
of the counterpart is fainter than $\approx 21$ mag. Its absolute magnitude 
would then be $M_{i^\prime} \gtrsim 9$ for a distance of 2.5~kpc.

This limit on the magnitude of the companion is not unreasonably faint 
for a polar. Eleven polars with orbital periods shorter than 2~h
are listed both with optical magnitudes in the catalog of \citet{rk03}, and 
with distances in \citep{war95}. We find that the absolute V-band magnitudes of
these 11 sources range from 11.3 -- 13.7 mag when they are in low optical
states, and 7.3 -- 9.5 in high optical states. To convert these $M_V$ 
magnitudes to $M_{i^\prime}$ values, we note that the infrared colors of 
magnetic CVs are similar to M dwarf stars \citep{hoa02}, so that  
$V - i^\prime \approx 2$ \citep[compare][]{fuk96,kil98}. Thus, 
we expect that in low optical states these polars would be 
comparable to or fainter than the lower limit of $M_{i^\prime} > 9$ mag 
for \forbit. 
However, all but one of these 
same polars (the exception being EP Dra) should be easily observable in 
their brighter optical states from 2.5 kpc away. Since the X-ray emission 
from \forbit\ clearly exhibits 
large variations in its X-ray intensity on time scales of a year
(Figure~\ref{fig:lcurve}), it probably also exhibits significant optical 
variability \citep{war95}. It is therefore possible that \forbit\ was 
observed in a low optical state in 2002 September.

\section{Conclusions}

We have identified eight sources of periodic X-ray variability in 
487~ks of observations of the 17\arcmin\ by 17\arcmin\ field centered on
the Galactic center that were taken over a two week span in 2002 May 22
through June 4 (Table~\ref{tab:per} and Figure~\ref{fig:per}). The X-ray
spectra of these sources indicate that seven lie near or beyond the 
Galactic center, while one is in the foreground at a distance of 
approximately 2.5~kpc (Table~\ref{tab:spec} and Figure~\ref{fig:spec}).
The spectra, luminosities, and periods of the Galactic center sources 
are all consistent with either magnetized CVs or HMXB pulsars; future
infrared observations will be required to determine the exact natures of
these sources.
The light curve of the foreground source indicates that it is an eclipsing
polar. We have searched for an optical counterpart in an $i^\prime$ image
obtained with the Clay (Magellan II) telescope, but none was found down
to a deabsorbed $i^\prime$ magnitude of 21. 
A polar in a low optical state at 2.5~kpc from Earth would probably have 
an $i^\prime$ magnitude
that is slightly fainter.

These sources are probably not unique to the 
Galactic center, as they are similar to a number of sources with hard 
spectra and X-ray periods longer than 100~s that recently have been 
identified in the plane of the Galaxy 
\citep{kin98, tor99, oos99, sug00, sak00}. 
However, they are the faintest sources ever observed to show periodicities.
The periodic variations were detectable only because \chandra\ observed
the Galactic center almost continuously for nearly 500~ks. Unfortunately, 
unless such long monitoring campaigns are carried out with \chandra\
in the future, the detection of similar periodicities is unlikely. The larger
effective areas of future X-ray missions may allow periodicities to be 
detected from sources using shorter observations. However, current plans 
for these missions lack 
the $< 1$\arcsec\ angular resolution required to resolve the faint X-ray 
sources in regions like the Galactic center, where both the density of 
sources and the flux from diffuse emission are high.

\acknowledgements{
We are grateful to S. Burles for obtaining the optical image of \forbit, 
and for advising us about how to process the image.
We also thank Z. Wang for advice on the optical processing, and M. Eracleous, 
D. Galloway, and J. Sokoloski for helpful discussions about 
the possible natures of these sources. We thank the referee, J. Grindlay, 
for his helpful comments. This work has been supported by 
NASA grants NAS 8-39073 and NAS 8-00128. W.N.B. also acknowledges the 
NSF CAREER award AST-9983783.
}

\begin{deluxetable}{llccccc}
\tablecolumns{7}
\tablewidth{0pc}
\tablecaption{Observations of the Inner 20 pc of the Galaxy\label{tab:obs}}
\tablehead{
\colhead{} & \colhead{} & \colhead{} & \colhead{} & 
\multicolumn{2}{c}{Aim Point} & \colhead{} \\
\colhead{No.} & \colhead{Start Time} & \colhead{ObsID} & \colhead{Exposure} & 
\colhead{RA} & \colhead{DEC} & \colhead{Roll} \\
\colhead{} & \colhead{(UT)} & \colhead{} & \colhead{(s)} 
& \multicolumn{2}{c}{(degrees J2000)} & \colhead{(degrees)}
} \startdata
1 & 1999 Sep 21 02:43:00 & 0242  & 40,872 & 266.41382 & $-$29.0130 & 268 \\
2 & 2000 Oct 26 18:15:11 & 1561 & 35,705 & 266.41344 & $-$29.0128 & 265 \\
3 & 2001 Jul 14 01:51:10 & 1561 & 13,504 & 266.41344 & $-$29.0128 & 265 \\
4 & 2002 Feb 19 14:27:32 & 2951  & 12,370 & 266.41867 & $-$29.0033 & 91 \\
5 & 2002 Mar 23 12:25:04 & 2952  & 11,859 & 266.41897 & $-$29.0034 & 88 \\
6 & 2002 Apr 19 10:39:01 & 2953  & 11,632 & 266.41923 & $-$29.0034 & 85 \\
7 & 2002 May 07 09:25:07 & 2954  & 12,455 & 266.41938 & $-$29.0037 & 82 \\[5pt]
8 & 2002 May 22 22:59:15 & 2943\tablenotemark{a}  & 34,651 & 266.41991 & $-$29.0041 & 76 \\
9 & 2002 May 24 11:50:13 & 3663\tablenotemark{a} & 37,959 & 266.41993 & $-$29.0041 & 76 \\
10 & 2002 May 25 15:16:03 & 3392\tablenotemark{a}  & 166,690 & 266.41992 & $-$29.0041 & 76 \\
11 & 2002 May 28 05:34:44 & 3393\tablenotemark{a}  & 158,026 & 266.41992 & $-$29.0041 & 76 \\
12 & 2002 Jun 03 01:24:37 & 3665\tablenotemark{a}  & 89,928 & 266.41992 & $-$29.0041 & 76 \\
\enddata
\tablenotetext{a}{These observations were used to search for sources with 
periodic variability.}
\end{deluxetable}

\begin{deluxetable}{lccccc}
\tablecolumns{56}
\tablewidth{0pc}
\tablecaption{Sources with Periodic X-ray Variations\label{tab:per}} 
\tablehead{
\colhead{Name} & \colhead{$N_{\rm tot}$} & \colhead{$B$} & \colhead{$P$} & 
\colhead{$A$} & \colhead{$T_0$\tablenotemark{a}} \\
\colhead{(CXOGC~J)} & \multicolumn{2}{c}{(counts)} & \colhead{(s)} & \colhead{(rms \%)} & 
\colhead{(MJD [TDB])} \\
} 
\startdata
174517.4$-$290650 & 273\tablenotemark{b} & 53.6 & 321.51(5) & 55(8) & 52416.9585(2) \\
174531.7$-$290542 & 233 & 67.6 & 16772(16) & 60(9) & 52416.981(7) \\
174532.3$-$290251 & 304 & 24.5 & 5612(2) & 58(6) & 52416.977(2) \\
174532.7$-$290552 & 1260 & 64.1 & 5409.6(6) & 71(2) & 52416.9409(7) \\ 
174534.5$-$290201 & 724 & 31.3 & 971.64(5) & 40(4) & 52416.9595(3) \\
174535.6$-$290034 & 466 & 70.3 & 9907(8) & 49(5) & 52417.011(4) \\
174541.8$-$290037 & 493 & 138.6 & 1092.15(6) & 53(6) & 52416.9546(4) \\
174543.4$-$285841 & 410 & 40.8 & 5313(2) & 48(6) & 52416.953(2) \\
\enddata
\tablenotetext{a}{Defined by zero phase when fitting a sinusoid to the 
fundamental frequency.}
\tablenotetext{b}{Only data from ObsIDs 2943 and 3663 were used for this 
source, as it later declined in flux and the pulsations disappeared.}
\tablecomments{1 $\sigma$ uncertainties on the last significant digit are 
indicated in parenthesis.}
\end{deluxetable}




\begin{deluxetable}{lcccccccc}
\tablecolumns{9}
\tablewidth{0pc}
\tablecaption{X-ray Spectra of Sources with Periodicities\label{tab:spec}} 
\tablehead{
\colhead{Name (CXOGC~J)} & \colhead{Flux} & \colhead{$N_{\rm H}$} &
\colhead{$\Gamma$} & \colhead{$E_{\rm Fe}$} & \colhead{$W_{\rm Fe}$} &
\colhead{$N_{\rm Fe}$} & \colhead{EW} & \colhead{$\chi^2/\nu$}
} 
\startdata
174517.4$-$290650 & 4.4 & $6^{+7}_{-2}$ & $1.0^{+1.7}_{-0.7}$ & $6.6^{+0.1}_{-0.1}$ & $0.2^{+0.1}_{-0.1}$ & $0.9^{+0.4}_{-0.4}$ & 960 & 63/42 \\
174531.7$-$290542 & 1.7 & $9^{+9}_{-7}$ & $-0.6^{+1.6}_{-1.3}$ & \nodata & \nodata & \nodata & \nodata & 13/11 \\
174532.3$-$290251 & 2.8 & $25^{+9}_{-10}$ & $1.8^{+1.5}_{-1.2}$ & $6.66^{+0.04}_{-0.03}$ & $<0.2$ & $0.8^{+0.5}_{-0.3}$ & 640 & 15/15 \\
174532.7$-$290552 & 3.2\tablenotemark{a} & $0.5^{+0.1}_{-0.1}$ & $1.4^{+0.1}_{-0.1}$ & \nodata & \nodata & \nodata & \nodata & 59/61 \\
174534.5$-$290201 & 6.8 & $13^{+8}_{-3}$ & $0.5^{+1.4}_{-0.5}$ & $6.6^{+0.2}_{-0.1}$ & $0.3^{+0.1}_{-0.1}$ & $1.7^{+1.4}_{-0.6}$ & 720 & 30/40 \\
174535.6$-$290034 & 3.3 & $13^{+5}_{-4}$ & $0.9^{+0.9}_{-0.8}$ & \nodata & \nodata & \nodata & \nodata & 25/25 \\
174541.8$-$290037 & 4.6 & $12^{+14}_{-7}$ & $0.3^{+1.0}_{-1.2}$ & $6.6^{+0.1}_{-0.1}$ & $0.13^{+0.11}_{-0.09}$ & $1.2^{+1.1}_{-0.4}$ & 920 & 28/28 \\
174543.4$-$285841 & 4.0 & $15^{+10}_{-3}$ & $0.6^{+1.4}_{-0.3}$ & \nodata & \nodata & \nodata & \nodata & 26/22 \\
\enddata
\tablenotetext{a}{The 0.5--8 keV flux for this foreground source was 
$3.7\times10^{-14}$ \ergcms.}
\tablecomments{The flux is in units of $10^{-14}$ \ergcms (2--8 keV). 
The column density 
$N_{\rm H}$ is in units of $10^{22}$ cm$^{-2}$. $\Gamma$ is the photon index.
The energy $E_{\rm Fe}$ and width $W_{\rm Fe}$ of the line emission 
are in keV, and the flux in the line $N_{\rm Fe}$ is in units of 
$10^{-6}$ \phcms. The equivalent width is in units of eV. 
Uncertainties are 90\% confidence ranges for one parameter
of interest.}
\end{deluxetable}



\begin{thebibliography}{0}
\bibitem[Apparao \etal(1994)]{app94} Apparao, K. M. V. 1994, {\it SSRev}, 
  69, 255
\bibitem[Baganoff \etal(2003)]{bag03} Baganoff, F. K. \etal\ 2003, \apj, 
  to appear in v590, n2
\bibitem[Broos \etal(2002)]{bro02} Broos, P., Townsley, L., Getman, K. \&
  Bauer, F. 2002, ``ACIS Extract, An ACIS Point Source Extraction Package'', 
  Pennsylvania State University, 
  http://www.astro.psu/xray/docs/TARA/ae\_users\_guide.html
\bibitem[Buccheri \etal(1983)]{buc83} Buccheri, R. \etal\ 1983, \aap, 128, 245
\bibitem[Campana \etal(2001)]{cam01} Campana, S., Gastaldello, F., Stella, L.,
  Israel, G. L., Colpi, M., Pizzolato, F., Orlandini, M., \& Dal Fiume, D.
  2001, \apj, 561, 924
\bibitem[Edmonds \etal(2003)]{edm03} Edmonds, P. D., Gilliland, R. L., Heinke, 
  C. O. \& Grindlay, J. E. 2003, \apj\ submitted, astro-ph/0307189
\bibitem[{Ezuka} \& {Ishida}(1999)]{ei99} {Ezuka}, H. \& {Ishida}, M. 1999,
  \apjs, 120, 277
\bibitem[Fukugita \etal(1996)]{fuk96} Fukugita, M., Ichikawa, T., Gunn, J. E.,
  Doi, M., Shimasaku, K., \& Schneider, D. P. 1996, \aj, 111, 1748
\bibitem[Garmire \etal(2002)]{gar02} Garmire, G. P., Bautz, M. W., 
  Ford, P. G., Nousek, J. A., \& Ricker, G. R. 2002, SPIE Vol. 4851, 
  ``Advanced CCD Imaging Spectrometer (ACIS) Instrument on the \chandra\ X-ray
  Observatory''
\bibitem[Gehrels(1986)]{geh86} Gehrels, N. 1986, \apj, 303, 336
\bibitem[Grindlay \etal(2001)]{gri01} Grindlay, J. E., Heinke, C., Edmonds, 
  P. D. \& Murray, S. S. 2001, {\it Science}, 292, 2290
\bibitem[Haberl \etal(1998)]{hab98} Haberl, F., Angelini, L., Motch, C., \&
  White, N. E. 1998, \aap, 330, 189
\bibitem[Hall \etal(2000)]{hal00} Hall, T. A., Finley, J. P. Corbet, R. H. D.,
  \& Thomas, R. C. 2000, \apj, 536, 450
\bibitem[Hellier \etal(1998)]{hmo98} Hellier, C., Mukai, K., 
 \& Osborne, J. P. 1998, \mnras, 297, 526
\bibitem[Hoard \etal(2002)]{hoa02} Hoard, D. W., Wachter, S., Clark, L. L., 
  \& Bowers, T. P. 2002, \apj, 569, 1037
\bibitem[Howell \etal(2001)]{hnr01} Howell, S. B., Nelson, 
  L. A., \& Rappaport, S. 2001, \apj, 550, 897
\bibitem[Iben, Tutukov, \& Fedorva(1997)]{itf97} Iben, I. Jr., Tutukov, A. V., 
  \& Fedorova, A. V. 1997, \apj, 486, 955
\bibitem[Israel \etal(2000)]{isr00} Israel, G. L. \etal\ 2000, \mnras, 314, 87
\bibitem[Kalogera(1999)]{kal99} Kalogera, V. 1999, \apj, 521, 723
\bibitem[Kent \etal(1991)]{kdf91} Kent, S. M., Dame, T. M., \& Fazio, G. 
  1991, \apj, 378, 131
\bibitem[Kilkenny \etal(1998)]{kil98} Kilkenny, D., van Wyk, F., Roberts, G.,
  Marang, F., \& Cooper, D. 1998, \mnras, 294, 93x
\bibitem[{Kinugasa} et al.(1998)]{kin98} {Kinugasa}, K. \etal\ 1998, \apj, 
  495, 435
\bibitem[Leahy \etal(1983)]{lea83} Leahy, D. A., Darbro, W., Elsner, R. F.,
        Weisskopf, M. C., Sutherland, P. G., Kahn, S., \& Grindlay, J. E.
        1983, \apj, 266, 160
\bibitem[Liu \etal(2000)]{liu00} Liu, Q.~Z., van Paradijs, J., 
  \& van den Heuvel, E.~P.~J. 2000, \aaps, 147, 25
\bibitem[Morris(1993)]{mor93} Morris, M. 1993, \apj, 408, 496
\bibitem[Muno \etal(2003)]{mun03} Muno, M. P., Baganoff, F. K., Bautz, M. W., 
  Ricker, G. R., Morris, M., Garmire, G. P., Feigelson, E. D., Brandt, W. N., 
  Townsley, L. K., \& Broos, P. S. 2003, \apj, 589, 225
\bibitem[Negueruela \etal(2000)]{neg00} Negueruela, I., Reig, P., 
  Finger, M. H., \& Roche, P. 2000, \aap, 356, 1003
\bibitem[Norton \& Watson(1989)]{nw89} Norton, A. J. \& Watson, M. G. 1989,
  \mnras, 237, 853
\bibitem[{Oosterbroek} et al.(1999)]{oos99} {Oosterbroek}, T., {Orlandini}, 
  M., {Parmar}, A.~N., {Angelini}, L., {Israel},
  G.~L., {Dal Fiume}, D., {Mereghetti}, S., {Santangelo}, A., \& {Cusumano}, G.
  1999, \aap, 351, L33
\bibitem[{Pfahl} et al.(2002)]{prr02} {Pfahl}, E., {Rappaport}, S., \& 
  {Podsiadlowski}, P. 2002, \apjl, 571, L37
\bibitem[Podsiadlowski, Rappaport, \& Pfahl(2002)]{prp02} Podsiadlowski, P., 
  Rappaport, S., \& Pfahl, E. D. 2002, \apj, 565, 1107
\bibitem[Pooley \etal(2002)]{pool02} Pooley, D. \etal\ 2002, \apj, 569, 405
\bibitem[Predhel \& Schmitt(1995)]{ps95} Predehl, P. \& Schmitt, 
  J. H. M. M. 1995, \aap, 293, 889
\bibitem[Predehl \& Truemper(1995)]{pt94} Predehl, P. \& Truemper, J. 1994, 
  \aap, 290, L29
\bibitem[Press \etal\ (1992)]{pre92} Press, W. H., Flannery, B. P., 
        Teukolsky, S. A., \& Vetterling, W. T. 1992, Numerical Recipes
        in C, 2nd Ed. (Cambridge: Cambridge University Press)
\bibitem[Ramsay \& Cropper(2003)]{rc03} Ramsay, G. \& Cropper, M. 2003, 
  \mnras, 338, 219
\bibitem[Rappaport \& van den Heuvel(1982)]{rap82} Rappaport, S. \& 
  van den Heuvel, E. P. J. 1982, IAU Symposium, 98, 327
\bibitem[Reig \& Roche(1999)]{rr99} Reig, P. \& Roche, P. 1999, \mnras, 
  306, 100
\bibitem[Rieke \& Lebofsky(1985)]{rl85} Rieke, G. H. \& Lebofsky, M. J. 1985,
  \apj, 288, 618
\bibitem[Ritter \& Kolb(2003)]{rk03} Ritter, H. \& Kolb, U. 2003, to 
  appear in \aap, astro-ph/0301444
\bibitem[{Sakano} et al.(2000)]{sak00} {Sakano}, M., {Torii}, K., {Koyama}, K.,
  {Maeda}, Y., \& {Yamauchi}, S. 2000, \pasj, 52, 1141
\bibitem[Schwarz \etal(2002)]{sch02} Schwarz, R., Greiner, J., Tovmassian, 
  G. H., Zharikov, S. V., \& Wenzel, W. 2002, \aap, 392, 505
\bibitem[Serabyn \& Morris(1996)]{sm96} Serabyn, E. \& Morris, M. 1996, \nat, 
  382, 602
\bibitem[Sugizaki \etal(2000)]{sug00} Sugizaki, M., Kinugasa, K., Matsuzaki, 
  K., Terada, Y., Yamauchi, S., \& Yokogawa, J. 2000, \apj, 534, L181  
\bibitem[{Torii} et al.(1999)]{tor99} {Torii}, K., {Sugizaki}, M., 
  {Kohmura}, T., {Endo}, T., \& {Nagase}, F. 1999, \apj, 523, L65
\bibitem[Townsley \etal(2000)]{tow00} Townsley, L. K., Broos, P. S.,
  Garmire, G. P., \& Nousek, J. A. 2000, \apj, 534, L139
\bibitem[Vaughan et al.(1994)]{vau94} Vaughan, B. A. et al. 1994, \apj, 
	435, 362
\bibitem[Verbunt \etal(1997)]{ver97} Verbunt, F., Bunk, W. H., Ritter, H., \&
  Pfeffermann, E. 1997, \aap, 327, 602
\bibitem[{Warner}(1995)]{war95} {Warner}, B. 1995, 
  {\em {Cataclysmic Variable Stars}}, Cambridge University Press
\bibitem[White \etal(1995)]{wnp95} {White}, N.~E., {Nagase}, F., 
  \& {Parmar}, A.~N. (1995). In {{Lewin}, W.~H.~G. and {van Paradijs}, J. 
  and {van den Heuvel},
  E.~P.~J.}, editor, {\em X-ray Binaries}, Cambridge University
  Press, pg. 1
\end{thebibliography}
\end{document}